# Magnetic Properties of Ni-Fe Nanowire Arrays: Effect of Template Material and Deposition Conditions

Shyam Aravamudhan[1*a)], John Singleton[2], Paul A. Goddard[3], Shekhar Bhansali[1,b)]

[1]Department of Electrical Engineering, Nanomaterials and Nanomanufacturing Research Center (NNRC), University of South Florida, Tampa, FL 33620
[2]National High Magnetic Field Laboratories (NHMFL), Los Alamos National Laboratory, Los Alamos, NM
[3]Clarendon Laboratory, Department of Physics, University of Oxford, Oxford, United Kingdom

[a)]Email: saravamu@gatech.edu, [b)]Email: bhansali@eng.usf.edu, Tel: 1-813-874-3593, Fax: 1-813-974-5250.

[*]Now with Microelectronics Research Center (MiRC), Georgia Institute of Technology, Atlanta, GA 30332.

## Abstract

The objective of this work is to study the magnetic properties of arrays of Ni-Fe nanowires electrodeposited in different template materials such as porous silicon, polycarbonate and alumina. Magnetic properties were studied as a function of template material, applied magnetic field (parallel and perpendicular) during deposition, wire length, as well as magnetic field orientation during measurement. The results show that application of magnetic field during deposition strongly influences the c-axis preferred orientation growth of Ni-Fe nanowires. The samples with magnetic field perpendicular to template plane during deposition exhibits strong perpendicular anisotropy with greatly enhanced coercivity and squareness ratio, particularly in Ni-Fe nanowires deposited in polycarbonate templates. In case of polycarbonate template, as magnetic field during deposition increases, both coercivity and squareness ratio also increase. The wire length dependence was also measured for polycarbonate templates. As wire length increases, coercivity and squareness ratio decrease, but saturation field increases. Such magnetic behavior (dependence on template material, magnetic field, wire length) can be qualitatively explained by preferential growth phenomena, dipolar interactions among nanowires, and perpendicular shape anisotropy in individual nanowires.

## I. INTRODUCTION

In recent years the increasing interest in highly-ordered artificial magnetic nanostructures has been driven not only by a desire to understand the fundamental properties of these materials but also by the diversity of their potential applications. Such applications range from magnetic recording to sensors and to bio-magnetism.[1-3] Nanoscale magnetic arrays are attractive as ultra-high density storage media. The magnetic density in conventional longitudinal recording is typically less than 50 Gb/in$^2$, limited by thermal instability.[35] However, nanoscale magnetic arrays have the potential to produce recording up to 100 times greater than existing random access memories.[4-6] The other field of extremely promising applications is bio-magnetism, as the magnetic nanowires can be manipulated and probed by magnetic interactions.[7] The spectrum of applications in bio-magnetism includes cell separation,[9] bio-sensing,[10] cellular studies,[11,12] and a variety of other therapeutic applications.[7,8] Holmgren et al.[13] performed both high yield (> 90%) and high purity single step cell separations on NIH-3T3 mouse fibroblast cells by applying magnetic forces through nanowires. These nanostructures have also been explored for use in drug delivery and gene therapy. Further, as nanowires are quasi-one-dimensional, high-aspect ratio (> 100) structures they have a large surface to volume ratio. Thus, nanowire-based sensors allow for higher sensitivity, higher capture efficiency and faster response time, due to their large adsorption surface and small diffusion time.[14] Many types of magnetic nanowire arrays (metals, alloys, or multi-layer structures) have been previously investigated.[15-17] Amongst the various materials studied, Ni-Fe is attractive because of its superior ferromagnetic properties, high magnetization behavior and invar effect in certain compositions.[18,19]

One-dimensional nanostructures can be produced by a variety of techniques such as molecular beam epitaxy, nanolithography, vapor-liquid-solid growth, and electrodeposition. Electrodeposition of metals into the pores of nanoscale templates (such as alumina membranes, nuclear track-etched polymer membranes, mesoporous silica or porous silicon) has been particularly attractive [20,21] because: (a) it is a simple, low-cost, high-throughput technique for fabricating large arrays of nanowires with monodispersive diameter and length; (b) it provides the ability to tailor size, length, shape and morphology of the material deposited by controlling the template morphology and the synthesis parameters; and (c) it provides the ability to introduce composition modulation along the wire length, which in turn enables precise control on architecture and magnetic properties. For example, Reich et al.[1] showed selective binding of two different ligands onto two-component Ni/Au nanowires, thus enabling spatially modulated functionalization schemes. Such properties can potentially give rise to improved performance in bio-magnetic applications.

To date, most of the research work has focused on studying the magnetic properties by changing the electrodeposition parameters or template parameters such as pore diameter, inter-pore spacing.[22, 23] In this work, we present a comparative study of magnetic nanowires deposited in different templates. The magnetic properties of nanowire arrays are directly related to the template properties - pore dimensions, relative pore orientation, pore size distribution and pore surface roughness. In addition to traditional templates such as porous alumina and polycarbonate, magnetic properties of nanowires deposited in porous silicon are also investigated. In a previous work, we have demonstrated the ability to control the porous silicon dimensions (pore diameter, 40-290 nm and length, up to 240 µm) and then successfully electroplated metal ions into the pores.[24] In order to investigate the magnetostatic coupling effect on the overall magnetic properties, nanowires with different wire lengths were prepared by controlling the electroplating time. We also investigated the influence of applied magnetic field during electrodeposition of Ni-Fe nanowires on their crystallographic and magnetic properties.

## II. EXPERIMENTAL DETAILS

Magnetic nanowire arrays are prepared by electroplating Ni-Fe into the pores of Anopore® alumina membranes,[26] Nuclepore Polycarbonate track-etched membranes,[26] and in-house prepared porous silicon templates. The alumina and polycarbonate membranes are thoroughly cleaned in de-ionized water and subsequently dried prior to use. The porous silicon template is prepared in-house by electrochemical etching of silicon substrate.[24,25] N-type 2" silicon substrate (resistivity: 0.4-0.6 ohm-cm) is etched in a mixture of 1:1::49% HF:ethanol at a constant current density of 35 mA/cm$^2$. Ethanol is added to the HF solution to (a) increase the wettability of porous silicon surface and (b) to remove hydrogen evolved during etching.

A film of aluminum (~ 1 µm) is evaporated on one side of all templates to serve as the working electrode. This is followed by electrochemical deposition of Ni-Fe into the templates from a sulfate based electroplating bath (200g/L NiSO$_4$.6H$_2$O, 8g/L FeSO$_4$.6H$_2$O, 5g/L

NiCl$_2$.6H$_2$O, 25 g/L H$_3$BO$_3$, 3g/L Saccharin). As the stoichiometry of the Ni-Fe nanowires is significantly affected by plating temperature, pH, agitation conditions, current, and additives, all the parameters have been maintained constant, except for the electroplating time. In the polycarbonate template, varying lengths of nanowire are deposited (up to 5 µm) so that the length dependent magnetic properties can be studied. Ni foil is used as anode to maintain constant metal-ion composition and the electroplating is performed under a current density of 3 mA/cm$^2$ at room temperature (20 ºC). The applied magnetic field during electroplating can be oriented either perpendicular or parallel to the template plane. Note that throughout the paper the orientation of applied magnetic fields will be described with respect to the template plane rather than the nanowire axis. This is because, although in general the wires are perpendicular to the plane of the template, some samples exhibit a degree of misalignment. This point will be discussed in detail later.

The structure and morphology of the nanowires are analyzed under a Scanning Electron Microscope (SEM). Energy Dispersive Spectroscopy (EDS) and X-ray diffraction are used to investigate the composition and crystallographic structure of the electrodeposited Ni-Fe (after etching Al). The X-ray diffraction is performed using an X'Pert PRO X-diffraction system (XRD) from Philips Analytical with a monochromatized Cu Kα ($\lambda$ = 15.4 nm) radiation in a Bragg-Brentano arrangement. A Physical Property Measurement System (PPMS) and Superconducting Quantum Interference Device (SQUID) magnetometer from Quantum Design are used to measure the magnetic properties of the nanowires embedded within the templates.

### III. EXPERIMENTAL RESULTS AND DISCUSSION

#### A. Microscopy and Structural Characterization

Figure 1(a) shows an SEM micrograph of a cross-section of the silicon template after electrochemical etching for 100 minutes. The average pore diameter is 300±10 nm with an interpore distance of 850±50 nm. The pores are 145 μm in length and suffer from irregular walls and branching.[24] Figure 1(b) and 1(c) show the SEM micrographs of the top surface of the alumina and polycarbonate templates of thicknesses 60 μm and 6 μm respectively. The pores are 190±10 nm in diameter and have an interpore distance of 285±15 nm in the alumina and 520±125 nm in the polycarbonate template. Table 1 summarizes the template/pore characteristics, as measured with the SEM. It is evident from the table that the lattice parameter, or inter-pore spacing, of polycarbonate and porous silicon templates are much larger than that of alumina.

Figures 2(a) and 2(b) show the SEM images of Ni-Fe nanowires electrodeposited in the porous silicon, alumina and polycarbonate templates respectively after removal of the template. The nanowires deposited in porous silicon are found to be 275±25 nm in diameter, while those deposited in both the alumina and polycarbonate are 190±10 nm in diameter. The length of the nanowire, which is initially estimated from the deposition charge and time, is later verified using the SEM. The Ni-Fe nanowires deposited in porous silicon exhibit a textured and highly faceted wire surface with multiple grain boundaries, wire breakage and branched growth of wires. This is probably due to the nature of the porous silicon etching.[24,29] In contrast, the nanowires deposited in the alumina and polycarbonate templates have smooth and uniform surface morphology. However, those wires deposited in the commercially available polycarbonate templates are found have an angle between the wire/pore axis and the normal to the plane typically between 0º and 34º.[26-28]

A quantitative EDS spectrum was taken to determine the elemental composition of the Ni-Fe nanowires deposited in the silicon, alumina and polycarbonate templates and the results are shown in Table 2. EDS analysis demonstrates that the atomic ratios of Ni and Fe in the nanowires formed in porous silicon, alumina and polycarbonate templates are close to 77:13, 85:14 and 84:15 respectively. A small amount of oxygen is seen in all the spectra, indicating a modicum of absorption from air. Lower elemental composition of Ni-Fe in porous silicon template is due to formation Si impurity phases, whose crystallographic structure is investigated by the XRD technique.

Figure 3 shows the x-ray diffraction patterns from the Ni-Fe nanowires (a) deposited in silicon without magnetic field, and (b) deposited in polycarbonate in the presence of a magnetic field (320 Oe) applied perpendicular to the template plane during electrodeposition. Significant differences in crystalline structure are observed. The diffraction patterns further confirm the electrodeposition of Ni-Fe alloy along with pure Ni. In absence of magnetic field, the pattern (Figure 3a) shows a strong peak for (111) $FeNi_3$ and (111) Ni along with other lesser peaks at (200), (211) for $FeNi_3$ and Ni. There is also evidence for formation of Ni-Si impurity phases from the silicon template. A strong peak at (111) for Ni-Fe and Ni indicates grain orientation along the preferred (111) direction, but the other peaks suggest an overall polycrystalline nature. The peak for Al may be due the aluminum sample stage, in case of the porous silicon sample.[24] Similar XRD diffraction peak intensities are obtained with alumina and polycarbonate templates electrodeposited in the absence of magnetic field. Note that all the samples measured have the same mass of Ni-Fe electrodeposits.

When the nanowires are deposited with a perpendicular magnetic field (Figure 3b), along with Ni-Fe (111), Ni-Fe grains with (200) texture also become dominant. This indicates a difference in crystal structure for nanowires grown in polycarbonate with applied magnetic field. Furthermore, SEM images (figure 2c) shows significant morphological changes such smoother walls for this type of nanowires. This is in accordance with the earlier reported data of uniform morphology of the electrodeposited films obtained in applied magnetic field.[33] Importantly, the applied magnetic field seems to have enhanced the growth of Ni-Fe (200) textures compared to (200) textures, when no magnetic field was applied. As suggested by Devos et al.[30] and Tabakovic et al.,[33] this may be a consequence of induced convective solution flow due to magnetohydrodynamic effect near the template's vicinity. This is turn may cause decrease in the thickness of the diffusion layer and therefore an increase in the mass transport of active species. Furthermore, in the presence of applied magnetic field, enhanced Ni-Fe (200) texturing indicates towards forced growth of Ni-Fe grains with their *c*-axis parallel to the orientation of the applied field.[30] This result is consistent with published results for Co nanowire arrays.[28] It seems however, that the applied field is not strong enough to totally force the Ni-Fe *c*-axis to align perpendicular to the plane of the template. Therefore, Ni-Fe (111) textures do not disappear.

B. Magnetic Characterization

Next, we compare the magnetic properties of the Ni-Fe nanowires deposited in the three different templates (porous silicon, alumina and polycarbonate) with/without a magnetic field of 320 Oe applied perpendicular to the plane of the template during electrodeposition. The saturation magnetizations of Ni-Fe were measured to be about 1 T. Magnetization hysteresis loops, which display the magnetic response of a material to an external field have been used to characterize Ni-Fe nanowires. The hysteresis loops may generally depend on the material, size and shape, microstructure and the orientation of applied magnetic field with respect to the sample. In case of nanowires, the key dependent property is magnetic anisotropy, which is the sum of different contributing factors such as shape anisotropy, magnetocrystalline anisotropy, magnetostatic coupling and other morphological characteristics. It should be noted that in nanowires with no preferential orientation, the magnetocrystalline anisotropy can compete with the shape anisotropy. However, if the easy axis is aligned along the wire axis both shape and magnetocrystalline anisotropies will add up. Lastly, magnetostatic coupling will always reduce both coercivity and effective perpendicular magnetic anisotropy.[34]

Figure 4 depicts typical magnetic hysteresis curves of Ni-Fe nanowires deposited without magnetic field in (a) porous silicon, (b) alumina and (c) polycarbonate templates. The magnetization curves both parallel and perpendicular to the template plane are shown. Quasi-one-dimensional structures such as Ni-Fe nanowires might reasonably be expected to behave like infinitely long, magnetic cylinders. If so, they should exhibit strong anisotropy, with the magnetic easy-axis aligning parallel to the wires.[31] In addition, if the magnetocrystalline anisotropy is small compared to shape anisotropy, square hysteresis curves are expected when the magnetization is measured along the cylindrical axis.[31] However, it is clear from Figure 4 that these nanowires exhibit little or no magnetic anisotropy. This is in agreement with our earlier publication in which we suggested that (a) the presence of branched and rough wire surfaces may reduce the shape anisotropy term and (b) competition between this 'reduced' shape anisotropy and magnetocrystalline anisotropy (with no preferential orientation along the easy axis, as seen in figure 3a) may result in zero overall magnetic anisotropy.[24]

The coercivity and squareness ratio (defined as ratio of the remnant magnetization to the saturation magnetization) are in the range of 50-100 Oe and 0.1-0.18 respectively for all the samples deposited in the absence of a magnetic field (see Table 3). We note that a similarly weak magnetic anisotropy is shown by nanowires deposited in small magnetic field of 320 Oe applied parallel to the template plane.

A very different behavior is exhibited by the nanowires electrodeposited in the presence magnetic field of 320 Oe applied perpendicular to the template plane. Figure 5 shows the typical magnetic hysteresis curves for these samples in the three different templates. The coercivities and squareness ratios are tabulated in Table 3. It is seen that for all the samples, the magnetic anisotropy is enhanced compared to those deposited in zero field. In all cases, the coercivity and squareness ratio is larger for the magnetization measured perpendicular to the template plane (i.e. roughly parallel to the nanowire axis), but the precise value of these parameters depends on the template material.

Figure 5a and Table 3a show the data for wires deposited in the porous silicon template. Although there is enhancement of the perpendicular squareness ratio and coercivity when the sample is deposited in an applied field, this increase is smaller compared to other two templates. Given the discussion about wire morphology above and by Aravamudhan et al.,[24] it seems likely that further enhancement of these parameters is hindered by the imperfections and surface roughness inherent in nanowires formed using the silicon templates.

In case of the alumina template (figure 5b and table 3b), because of small magnetocrystalline anisotropy (both Ni-Fe (111) and (200) textures equally being dominant), the net magnetic anisotropy is mainly due to two terms: (a) shape anisotropy induced due to magnetic easy axis parallel to wire axis, (b) magnetostatic coupling between wires, which develops an easy axis perpendicular to wire axis. Because of higher pore density in alumina template (about $10^9$ pores/cm$^3$) compared to polycarbonate or porous silicon (less than $10^8$ pores/cm$^3$), the net contribution from dipole field (aligned perpendicular to wire) is to reduce the effective anisotropy field given by[22, 23]

$$H_k = 2\pi M_s - \frac{6.3 M_s r^2 L}{d^3},$$

(1)

where, $M_s$ is the saturation magnetization, r is the wire diameter, L is the length and d is the interpore distance. The squareness ratios and coercivities in this case (for 3 different samples) were, however, slightly improved to 0.25-0.28 and 200-220 Oe respectively from greater oriented growth of nanowires. Finally, in the case of polycarbonate, (figure 5c and table 3c), a remarkable perpendicular anisotropy is exhibited. It can be seen that the maximum squareness ratio of 0.58-0.60 and coercivity of 400-425 Oe were observed (from 3 samples) when the measuring magnetic field is perpendicular to template plane. This suggests that application of perpendicular magnetic field during electrodeposition in polycarbonate template results in highest perpendicular magnetic anisotropy. However, the slight shearing of the hysteresis curve is mainly due to the 34º (maximum) deviation between the pore axis and surface normal,[26-28] along with the dipole interactions between the wires (interpore distance is 520±125 nm). Even though the average interpore distance in polycarbonate is much larger compared to alumina, according to Maeda et al.,[32] wire interactions will still occur for spacings up to 1.5 µm. This dipole wire interaction tends to align perpendicular to wire axis, resulting in a decrease in both squareness ratio and coercivity.

Next, the effect of varying the magnetic field perpendicular (270-1060 Oe) to the polycarbonate template during electrodeposition process was investigated. Figure 6 and table 4 show the measured average coercivity and squareness ratio ($M_r/M_s$) as a function of perpendicular magnetic field during electrodeposition. With increase in applied magnetic field during electrodeposition both coercivity and squareness ratio increase significantly. Squareness ratio of about 0.76 was observed for perpendicular magnetic field of 1060 Oe, indicating greater Ni-Fe growth with c-axis parallel to nanowire axis and hence enhanced perpendicular magnetic anisotropy. However, this is still a lower squareness ratio than the expected theoretical values because of the above stated reasons.

Lastly, the length effect was examined by depositing Ni-Fe nanowires of varying lengths (2-5 µm) in polycarbonate template in presence of perpendicular magnetic field of 320 Oe during deposition. Figure 7 and table 5 show the measured average coercivity and squareness ratio as a function of wire length. For magnetic field applied perpendicular to template plane, as the wire length is increased, according to infinite long magnetic cylinders model,[31] the shape anisotropy should also increase. But our experiments show that both coercivity and squareness monotonically decreases, with increase in wire length. This may be caused by the length dependence of dipole interactions among wires, given as[22,23]

$$H_d = \frac{4.2 M_s r^2 L}{d^3},$$

(2)

where, $M_s$ is the saturation magnetization, r is the wire diameter, L is the length and d is the interpore distance. In addition, as wire length increases, saturation magnetization also increases.

## IV. CONCLUSIONS

In summary, in this work, a systematic investigation was performed to study the structural and magnetic properties of Ni-Fe nanowires as a function of (a) template material (porous silicon, alumina and polycarbonate), (b) applied magnetic field during electrodeposition (0-1060 Oe), (c) wire length (2-5 μm) and (d) field orientation (parallel/perpendicular to template plane) during measurement. The applied magnetic field during electrodeposition was shown to have strong influence on crystallographic and magnetic properties of Ni-Fe nanowires, in particular, in the case of polycarbonate template, Ni-Fe nanowires of diameter 190±10 nm and length 2 μm fabricated in polycarbonate template with 1060 Oe applied magnetic field showed the highest coercivity of 530 Oe and squareness ratio of 0.74. The application of magnetic field perpendicular to the template plane during deposition tends to force the Ni-Fe grains with c-axis along the orientation of applied field, thereby resulting in perpendicular shape anisotropy. Further, the influence of applied magnetic field strength and nanowire length on magnetic properties was also studied. It was shown that with increase in magnetic field during deposition both coercivity and squareness ratio increased significantly, while coercivity and squareness monotonically decreased, with increase in wire length because of the length dependency on dipole interactions. The promising aspect of this work was the ability to tailor the magnetic and structural properties of Ni-Fe nanowires by application of strong magnetic field during electrodeposition and by selection of template material. Optimization of fabrication process to create high-density, isolated and vertical ferromagnetic nanowire arrays comparable to the theoretical expectations (based on coherent rotation theory) for coercivity and squareness ratio is currently underway. This is a key requirement for applications in ultra-high density magnetic storage and bio-magnetics.

## ACKNOWLEDGEMENTS

This work was supported the National Science Foundation (NSF) NER award no. ECS-0403800. P.A.G. thanks the Glasstone Foundation for financial support.

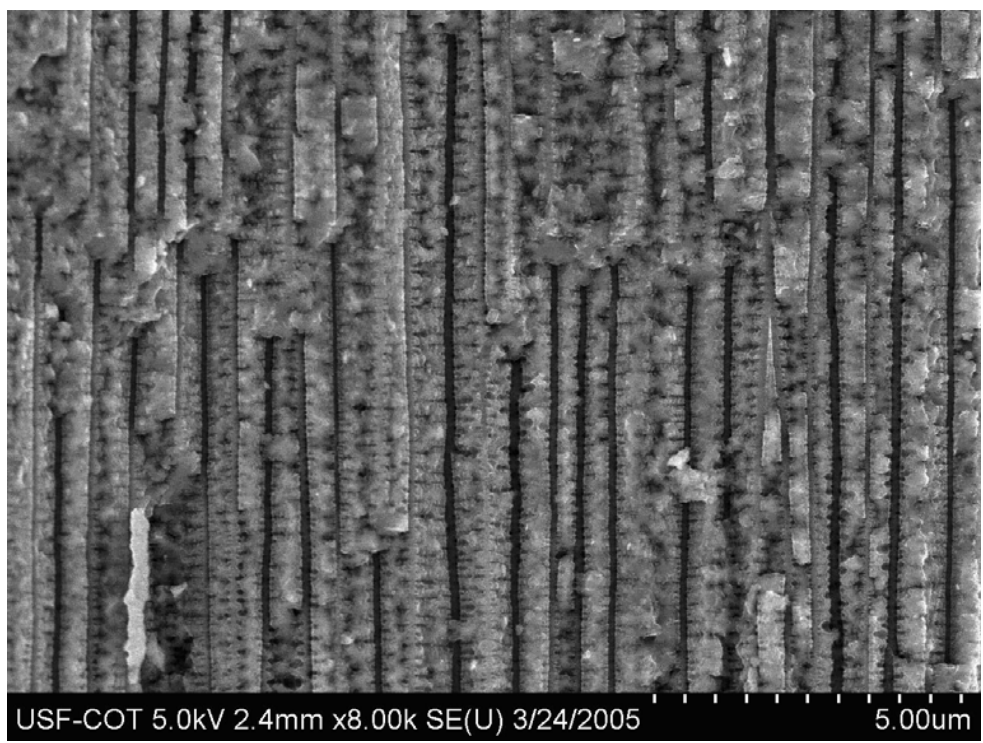

(a)

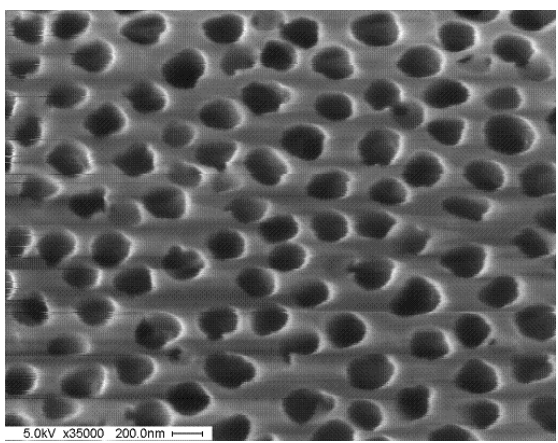

(b)

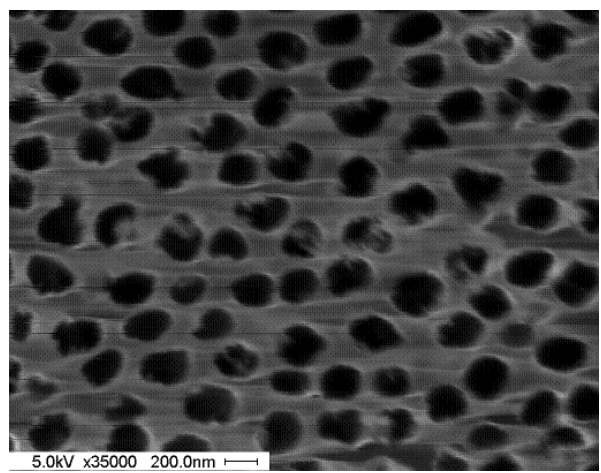

(c)

**Figure 1**: SEM images of the starting templates. **(a)** Cross-section of the n-type silicon substrate showing the 290±10 nm diameter and 145 μm deep nanopores created using electrochemical etching. View of the pores in **(b)** the alumina template (diameter = 200 nm, separation = 285±15 nm), and **(c)** the polycarbonate template (diameter = 200 nm, separation = 520±125 nm).

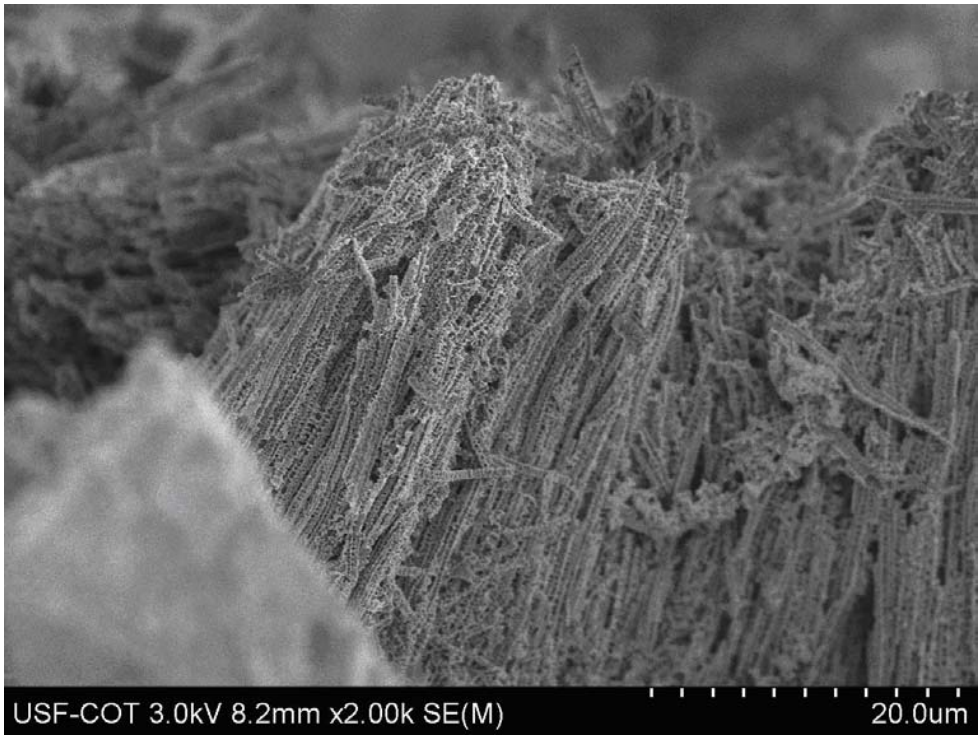

(a)

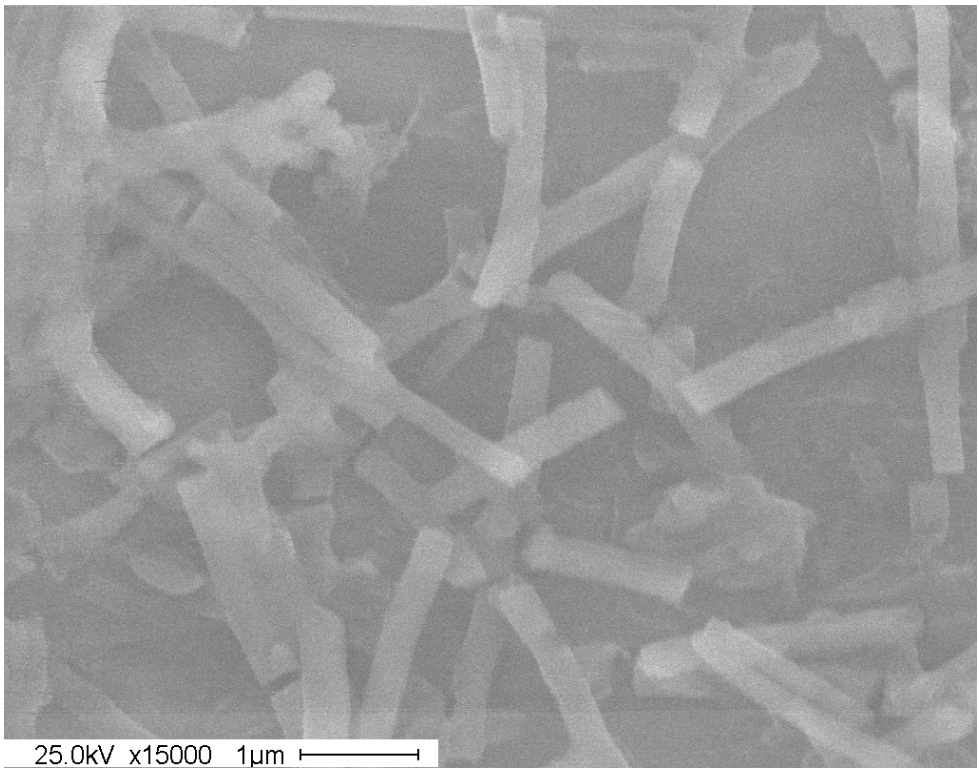

(b)

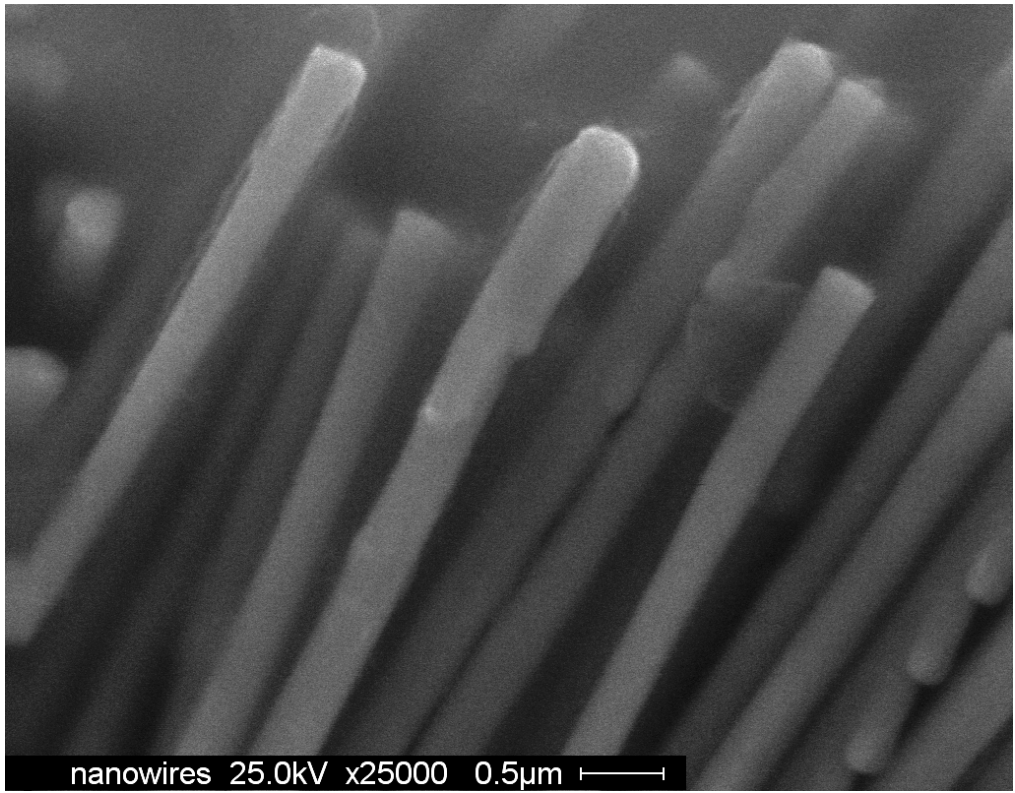

**(c)**

**Figure 2**: **(a)** SEM image of a cluster of 275±25 nm Ni-Fe nanowires released from the porous silicon template, **(b)** SEM view of released nanowires electrodeposited in perpendicular magnetic field in alumina template, **(c)** SEM view of released nanowires electrodeposited in perpendicular magnetic field in polycarbonate template. The alumina and polycarbonate nanowires wires are 190±10 nm in diameter, regular and uniform.

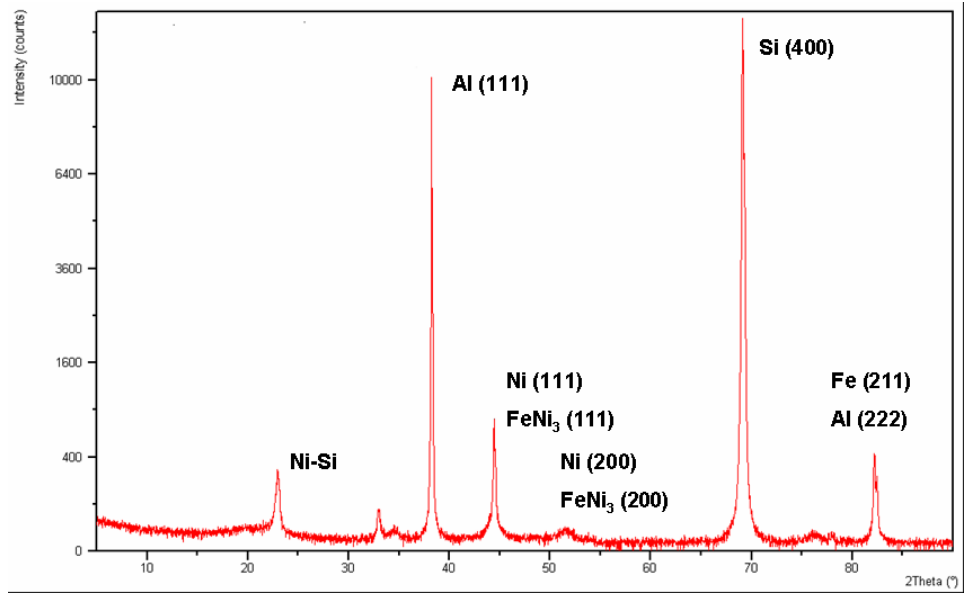

(a)

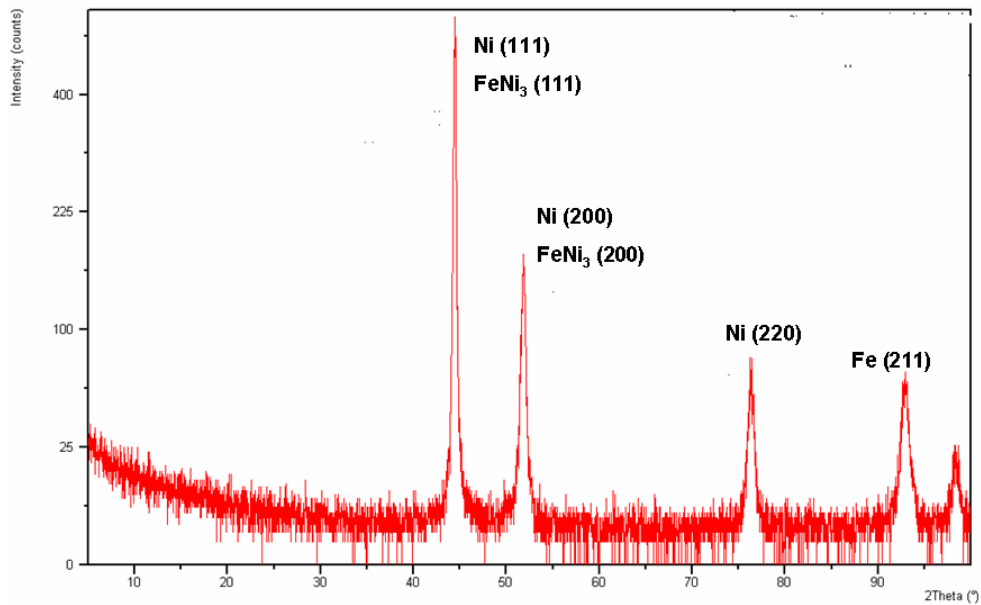

(b)

**Figure 3:** X-ray diffraction pattern of Ni-Fe nanowires deposited in: **(a)** porous silicon template with no applied magnetic field, **(b)** polycarbonate template with a magnetic field of 320 Oe applied perpendicular to the plane of the template during electrodeposition.

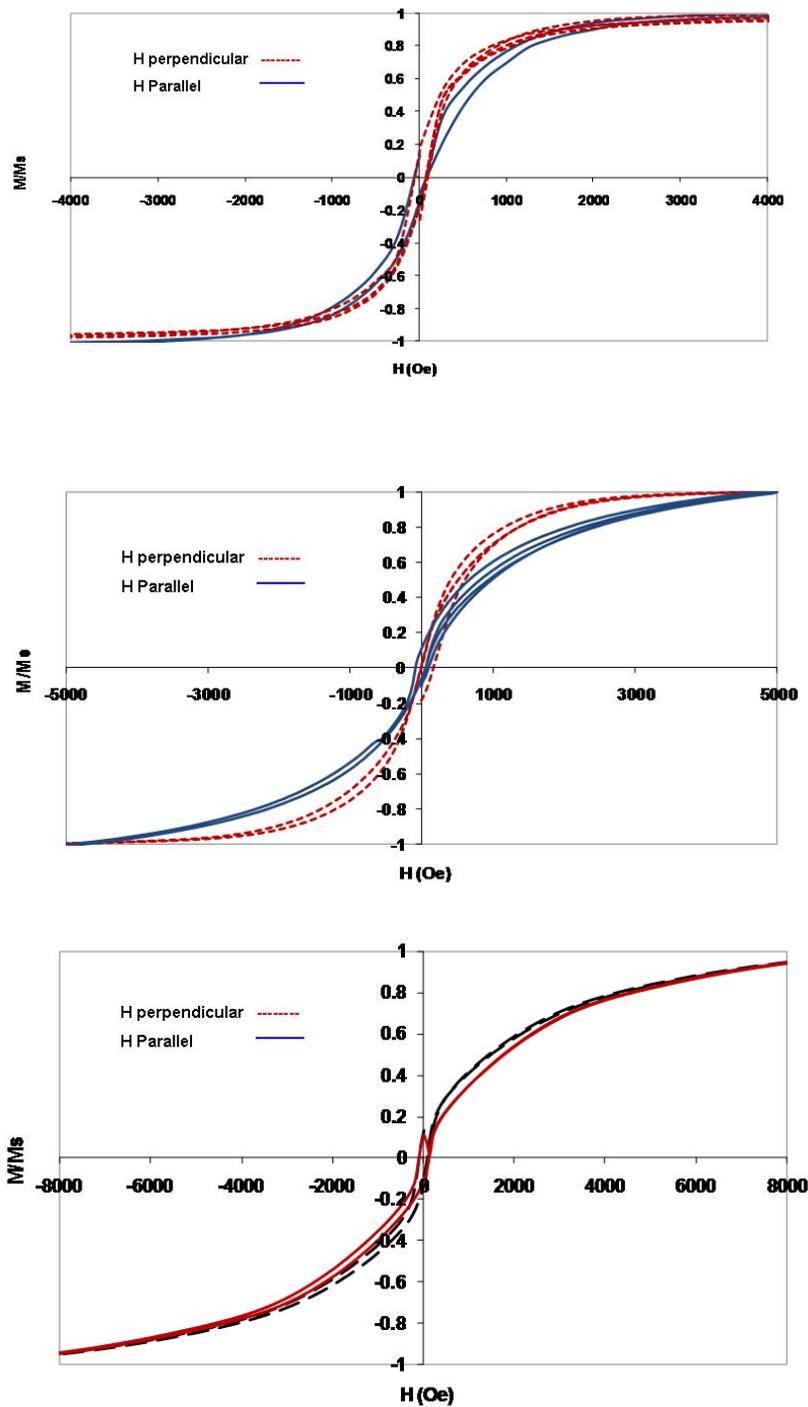

**Figure 4**: Typical magnetic hysteresis curves of Ni-Fe nanowire arrays electrodeposited in the absence of magnetic field in (**a**) porous silicon, (**b**) alumina, and (**c**) polycarbonate templates (average of 3 samples measured). Magnetization measured both parallel and perpendicular to the template plane are shown.

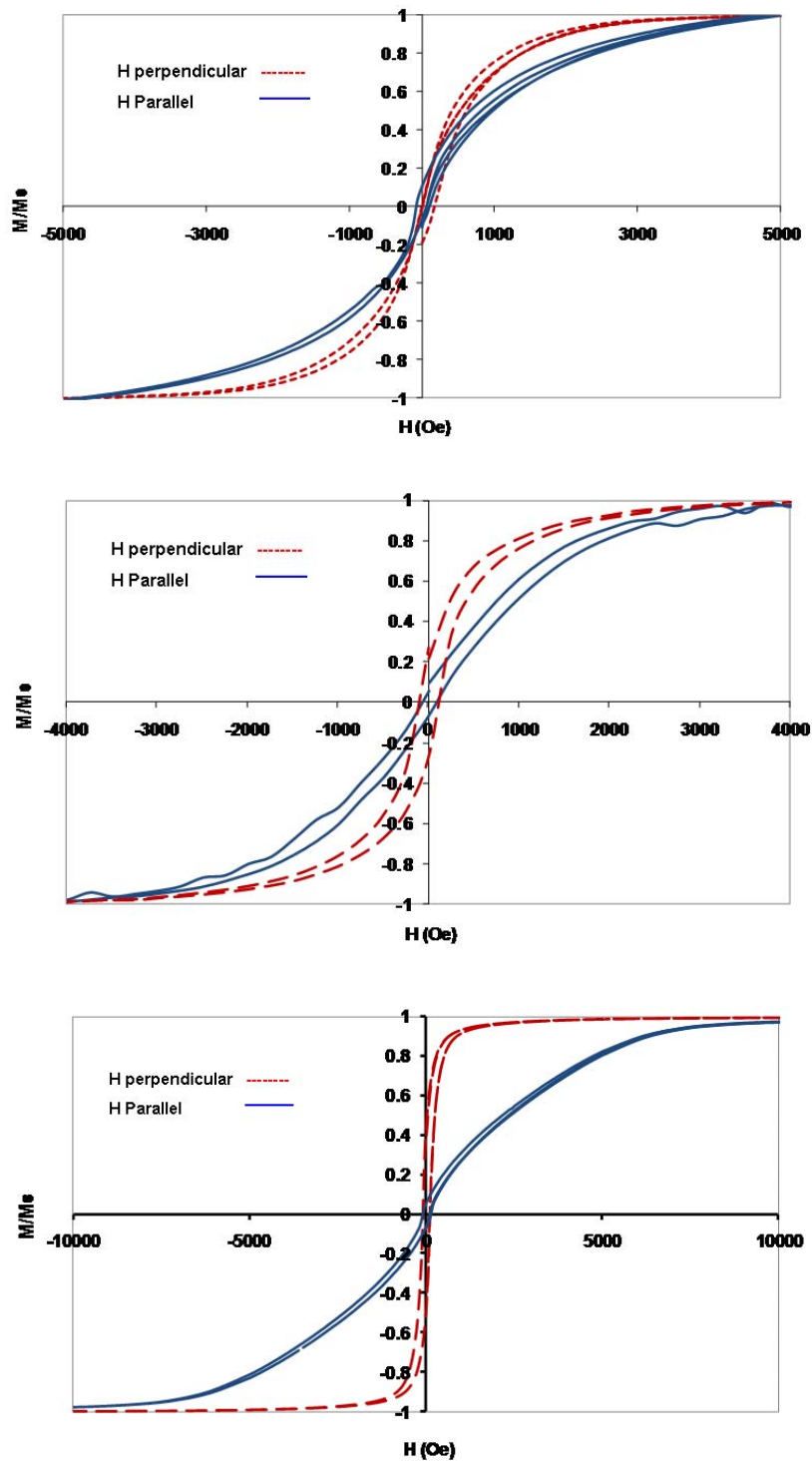

**Figure 5**: Typical magnetic hysteresis curves of Ni-Fe nanowire arrays electrodeposited in the presence of a small magnetic field (320 Oe) in **(a)** porous silicon, **(b)** alumina, and **(c)** polycarbonate templates (average of 3 samples measured). Magnetization measured both parallel and perpendicular to the template plane are shown.

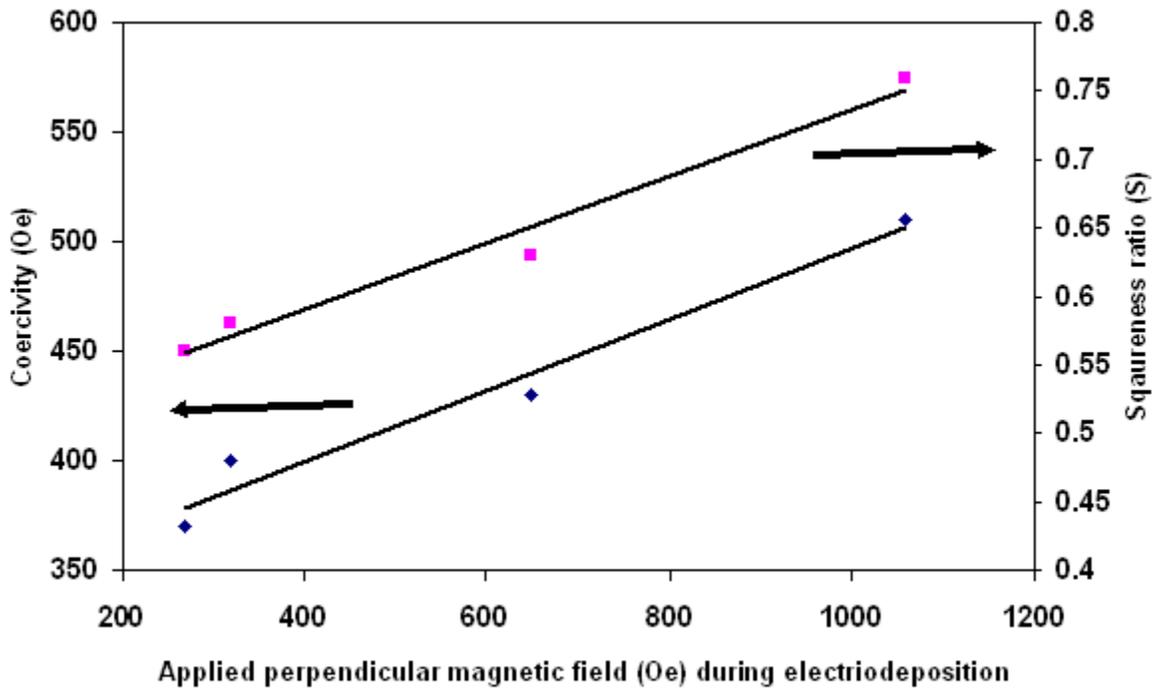

**Figure 6**: Dependence of coercivity and squareness ratio on applied perpendicular magnetic field during Ni-Fe electrodeposition in polycarbonate template.

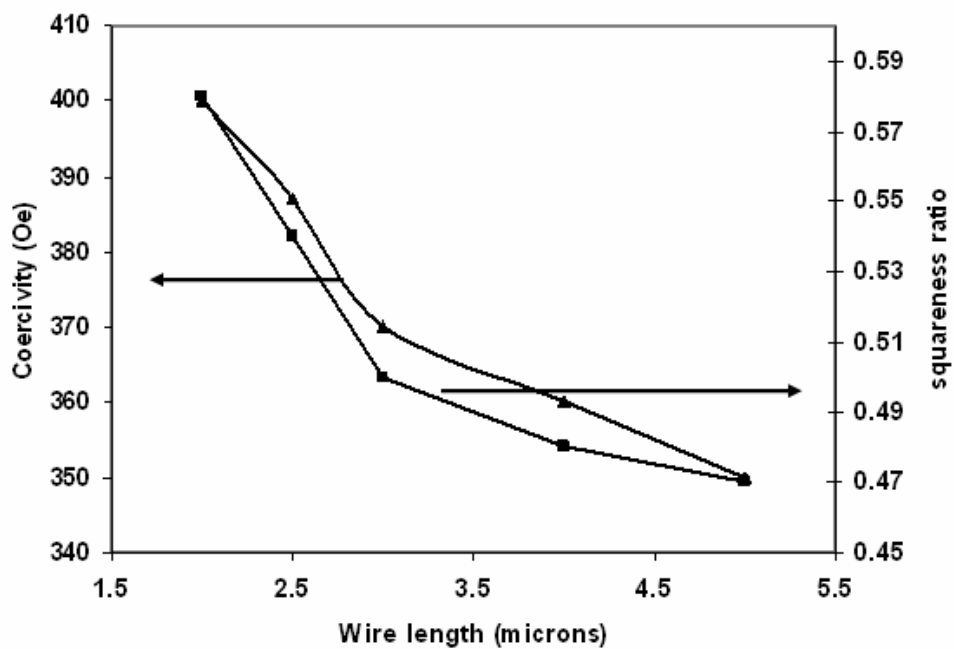

**Figure 7**: Dependence of coercivity and squareness ratio on Ni-Fe wire length, deposited in polycarbonate template.

**Table 1**: Template parameters and characteristics

| Parameters | Porous silicon | Polycarbonate | Alumina |
|---|---|---|---|
| Pore size (nm) | 300±10 | 190±10 | 190±10 |
| Inter-pore distance (nm) | 850±50 | 520±125 | 285±15 |
| Pore density (pores/cm$^2$) | about 10$^7$ | about 10$^8$ | about 10$^9$ |

**Table 2**: EDS elemental composition of Ni-Fe wires

| Ni-Fe Nanowires (Element) | In Porous Silicon At % | In Alumina At % | In Polycarbonate At % |
|---|---|---|---|
| O K | 9.77 | 1.21 | 1.03 |
| Fe K | 12.93 | 14.23 | 14.85 |
| Ni K | 77.30 | 84.56 | 84.12 |
| Total | 100.0 | 100.0 | 100.0 |

**Table 3**: Magnetic characterization parameters for Ni-Fe nanowires in different template and magnetic field during deposition (Data showed is from 3 samples each).

**(a)** Template: Porous Silicon

| Wire length (μm) | Magnetic field during deposition (Oe) | Coercivity (Oe) | Squareness ratio |
|---|---|---|---|
| 2 - 3 | 0 | 80-100 | 0.18 |
| 2 - 3 | 320 | 100-130 | 0.2-0.22 |

**(b)** Template: Alumina

| Wire length (μm) | Magnetic field during deposition (Oe) | Coercivity (Oe) | Squareness ratio |
|---|---|---|---|
| 2 - 2.3 | 0 | 60-80 | 0.15 |
| 2 - 2.3 | 320 | 200-220 | 0.25-0.28 |

**(c)** Template: Polycarbonate

| Wire length (μm) | Magnetic field during deposition (Oe) | Coercivity (Oe) | Squareness ratio |
|---|---|---|---|
| 2 - 2.2 | 0 | 50-65 | 0.12 |
| 2 - 2.2 | 320 | 400-425 | 0.58-0.60 |

**Table 4**: Magnetic characterization parameters for Ni-Fe nanowires of length 2-2.2 μm in polycarbonate template with varying magnetic field during deposition

| Magnetic field during deposition (Oe) | Coercivity (Oe) | Squareness ratio |
|---|---|---|
| 270 | 379 | 0.52 |
| 320 | 400-425 | 0.58-0.60 |
| 650 | 460 | 0.63 |
| 1060 | 530 | 0.74 |

**Table 5**: Magnetic characterization parameters for Ni-Fe nanowires of varying length in polycarbonate template with fixed magnetic field (320 Oe) during deposition

| Nanowire length (μm) | Coercivity (Oe) | Squareness ratio |
|---|---|---|
| 2-2.2 | 400-425 | 0.58-0.60 |
| 2.5-2.75 | 387 | 0.54 |
| 3-3.2 | 370 | 0.5 |
| 4-4.25 | 360 | 0.48 |
| 5-5.3 | 350 | 0.47 |